\documentclass[aps,preprint,twocolumn]{revtex4}
\usepackage{epsfig}
\usepackage{amsmath}
\usepackage{epsfig}
\usepackage{amssymb}

\newcommand{\specialcell}[2][c]{%
\begin{tabular}[#1]{@{}c@{}}#2\end{tabular}}
\begin{document}

\title
{\bf Three types of discrete  energy eigenvalues in complex PT-symmetric scattering potentials} 
\author{Zafar Ahmed$^{1,*}$, Sachin Kumar$^2$ and Dona Ghosh$^3$}
\affiliation{$~^1$Nuclear Physics Division,  Bhabha Atomic Research Centre, Mumbai 400 085, India\\
$~^*$Homi Bhabha National Institute, Mumbai 400 094 , India	\\
	 $~^2$Theoretical Physics Section, Bhabha Atomic Research Centre, Mumbai 400 085, India
	\\
	$~^3$Department of Mathematics, Jadavpur University, Kolkata 700032, India}
\email{1:zahmed@barc.gov.in, 2: sachinv@barc.gov.in,  3: rimidonaghosh@gmail.com,}   
\date{\today}

\begin{abstract}
For complex PT-symmetric scattering potentials (CPTSSPs) $V(x)= V_1 f_{even}(x) + iV_2 f_{odd}(x), f_{even}(\pm \infty) = 0 = f_{odd}(\pm \infty),  V_1,V_2 \in \Re $, we show that  complex $k$-poles of transmission amplitude $t(k)$ or zeros of $1/t(k)$ of the type $\pm k_1+ik_2, k_2\ge 0$ are physical which  yield three types of discrete energy eigenvalues of the potential.  These discrete energies are real negative, complex conjugate pair(s) of eigenvalues
(CCPEs: ${\cal E}_n \pm i \gamma_n$) and real positive energy called  spectral singularity (SS) at $E=E_*$ where the transmission and reflection co-efficient of $V(x)$ become infinite
for a special critical value of $V_2=V_*$. Based on four analytically solvable and  other numerically solved models, we conjecture that a parametrically fixed CPTSSP has at most one SS. When $V_1$ is fixed and $V_2$ is varied there may exist Kato's exceptional point(s) $(V_{EP})$ and  critical values  $V_{*m}, m=0,1,2,..$,  so when $V_2$ crosses one of these special values  a new CCPE is created. When $V_2$  equals a critical value $V_{*m}$  there exist one SS at $E=E_*$ along with $m$ or more number of CCPEs. Hence,  this  single positive energy $E_*$ is the upper (or rough upper)  bound to the  CCPEs: ${\cal E}_l  \lessapprox  E_*$, here ${\cal E}_l$ corresponds to the last  of CCPEs.  If $V(x)$ has Kato's  exceptional points (EPs: $V_{EP1}<V_{EP2}<V_{EP3}<...<V_{EPl}$), the smallest of critical values $V_{*m}$ is always larger than $V_{EPl}$. Hence, in a CPTSSP, real discrete eigenvalue(s) and the SS are mutually exclusive whereas CCPEs and the SS  can co-exist .	  
\end{abstract}
\maketitle

\section{Introduction}
 So far, much attention has not been paid to complex conjugate pair(s) of eigenvalues (CCPEs) in a complex PT-symmetric scattering potential (CPTSSP):
 \begin{eqnarray}
 V(x)= V_1 f_{even}(x)+ iV_2 f_{odd}(x),  \nonumber \\ f_{even}(\pm \infty)=0=f_{odd}(\pm \infty),  V_1,V_2 \in \Re.
   \end{eqnarray} 
Here $f_{even}(x)$ is also positive definite so when $V_1<0$, the real part is a potential well.
     Remarkably, a recent proposal of the splitting [1] of spectral singularity (SS) [2] in coherent perfect absorption (CPA)-Lasers [3] brings them in focus. Here,  we show that  complex $k$-poles of the transmission amplitude $t(k)$ or zeros of $1/t(k)$ of the type $\pm k_1+ ik_2, k_2>0$ yield three kinds of discrete eigenvalues ($E=k^2)$. When $k_1 \ne 0$, we get a finite number of  CCPEs (${\cal E}_n \pm i \gamma_n$) of bound state of $V(x)$. For the real discrete bound states eigenvalues $k_1=0$ and $k_2>0$. Further, in a parametric evolution when the strength $V_2$ of the imaginary part of the $V(x)$ (1) admits a special (critical) value $V_*$, $k_2$ vanishes  and $k=k_1=k_*$, there occurs  an spectral singularity  SS in $V(x)$ at $E=E_*=k_*^2$. Based on four analytically solvable and  other numerically solved models, we conjecture that for a potential (1) whose parameters are fixed $(V_1,V_2)$, if $E_*$ exists it is unique (single) and it is the upper (or rough upper)  bound to the  CCPEs such that ${\cal E}_l  \lessapprox  E_*$, here ${\cal E}_l$ corresponds to the last  of CCPEs.  We show that when $V_2 > V_*$ ($V(x)$ is made  slightly more non-Hermitian than the critical one), the SS disappears and the last of CCPEs appears as though SS is split in to  ${\cal E}\pm i \gamma$, where ${\cal E} \approx  E_*$. 
Complex PT-symmetric potentials [4] which are invariant under the joint action of Parity ($x\rightarrow -x)$ and Time-reversal ($i\rightarrow -i$) are now well known to have real discrete spectrum if the PT-symmetry is exact when the energy eigenstates are also eigenstates of PT, else the energy eigenvalues could be a mixture of negative real discrete and CCPEs, then the  PT-symmetry is said to be broken (inexact) [1]. This happens below/above a critical value of the real parameter of the potential. Scarf II, is the the first and the simplest exactly solvable model $[V_S(x,U_1,U_2)=-U_1 \mbox{sech}^2x+iU_2 \mbox{sech}x \tanh x], U_1>0$ which explicitly demonstrates a phase transition of eigenvalues from real to complex conjugate pairs when $U_2=U_c=U_1+1/4$ [5].
In terms of Kato's exceptional points (EPs), in the parametric evolution of eigenvalues $E_n(U_2)$, $U_2=U_1+1/4$ is the unique (single) EP of the non-Hermitian Hamiltonian:$ H=p^2+V_S(x,U_1,U_2)$, where pairs of real discrete eigenvalues coalesce. It would be well to remark that the CCPEs are finite and the corresponding eigenfunctions must satisfy the Dirichlet boundary condition: $\psi(\pm \infty)=0$.

Scattering from complex non-Hermitian potentials starting from  the non-reciprocity of reflection for the left/right incidence [6]  has been well developed as coherent injection of beams at optical mediums specially the PT-symmetric ones which are realized as having equal loss and gain. The crucial existence of SS has been revealed [2] wherein the reflection and transmission probabilities for a complex PT-symmetric potential become infinite at a real positive energy $E_*$. The most novel phenomena of coherent scattering from complex PT-symmetric potentials
is coherent perfect absorption with lasing (CPA-Laser), where in terms of two port scattering matrix $S(E)$, $|\det(S(E))|=1$ which becomes indeterminate ($0/0$) at $E=E_*$ such that $\lim _{E \rightarrow E_*} |\det(S(E))| \rightarrow 1 [3].$

A very interesting aspect of investigations in novel phenomena [7] of coherent scattering from non-Hermitian potential lies in their imperfections. Non-reciprocity of reflection has been stated and proved [6] if a potential is non-real, asymptotically converging to zero  and spatially non-symmetric, however there could be an exception to  it [8]. Spectral singularity was revealed [2] for complex PT-symmetric potentials, it turns out that it is actually a property of non-Hermitian potentials [10]. Any aggregate of matter was stated to perfectly absorb coherently injected  beams at it from left and right, provided a small dissipation (imaginary part) is added to its refractive index. Hence the phenomenon of coherent perfect absorption (CPA) or time-reversed laser has been invented [10], but it turns out that it is the property of non-Hermitian potentials [9] and CPTSSPs are exceptions to it. These potentials instead display CPA with lasing [3]. However, it has not been investigated so far, that whether SS and real discrete spectra are mutually exclusive in a  parametrically fixed CPTSSP. Importantly, the  discrete real positive energy SS found [11] in the complex PT-symmetric version of Scarf II [12] was noted [13] to be the last of CCPE of the potential and the corresponding eigenstate was shown to be just a plane wave [2], yet the phrase and the  discussion of spectral singularities (a plural term) in Scarf II in  Ref. [13] undermines the crucial  singleness of SS. Here, based on four model potentials, we conjecture that the SS in a parametrically fixed CPTSSP is unique (it occurs  only at one energy denoted as $ E_*$) and no exception to this exists so far.  

\section{Various discrete eigenvalues in a complex PT-symmetric potential}
 CPTSSP are enriched with three types of discrete energies, these are negative real, CCPE and SS. So far, CCPEs have not received much attention, however, after a recent proposal of splitting [1] of SDSS, their importance has been underlined.  Here, we argue the possible existence of three types of discrete eigenvalues in a complex PT-symmetric potential in a simple model independent way in terms of poles of $r(k)$ and $t(k)$. All $k$-poles of  these amplitudes are not physical as they may give rise to eigenstates which diverge asymptotically.

Let $V(x)$ (1) be a CPTSSP which vanishes asymptotically. For the left incidence of a particle we can write the asymptotic solution of Schr{\"o}dinger equation 
\begin{equation}
\frac{d^2\psi(x)}{dx^2}+\frac{2\mu}{\hbar^2}[E-V(x)]\psi(x)=0
\end{equation}
as
$\psi(x\sim -\infty)= A e^{ikx}+ B e^{-ikx}$ and $\psi(x\sim \infty)=C e^{ikx}$, the reflection and transmission amplitudes are $r=B(k)/A(k)$ and $t(k)=C(k)/A(k)$, respectively. Now  let $k=\pm k_1+ik_2$
and let $A(\pm k_1+ik_2)=0$ such that $r,t$ and the reflection $R(k)$ and transmission probabilities  $T(k)$ become infinity. In this situation, the asymptotic solutions of Schr{\"o}dinger equation (2) become bound states as  $\psi(x\sim -\infty)= B e^{\mp ik_1 x} e^{k_2 x}$ and $\psi(x\sim \infty)= C e^{\pm i k_1 x} e^{-k_2 x}$, which vanish asymptotically satisfying the Dirichlet boundary condition $\psi(\pm \infty)=0$. Such states  called bound states with complex conjugate pair of eigenvalues. In the complex $k$-plane, these $k$ values are poles of $r(k)$ and $t(k)$ which lie  symmetrically in the first and second quadrant. Whenever $k_2$ becomes zero, they represent plane waves and this energy $E=k_1^2=E_*$ is called SDSS. This real discrete energy is embedded in positive energy continuum at which the eigenfunctions become  plane waves at both ends of the potential. The real discrete spectra can be visualized as those poles of $r$ and $t$ where $k_1=0$ and $k_2>0$, so in complex $k$-plane these are poles  which lie on the positive y-axis. 

In  1900, Planck prophesied that in microscopic world energies are quantized,  meaning energy  admits only special discrete values. Later, Sommerfeld (1916) justified the discrete energies as due to the quantization of phase-space. It was in 1926 that Schr{\"o}dinger
revealed that wave function $\psi(x)$ vanishes asymptotically  at these discrete energies and $\psi(x)$ is $L^2$ integrable. In 1928,  Gamow revealed the resonances to be actually complex discrete eigenvalues ($E_n-i\Gamma_n/2$) at which $\psi_n(x,t)$ grows spatially at asymptotic distances and decays time-wise. According to this, one has to impose the condition of outgoing wave boundary condition at the exit of the potential to extract possible discrete complex energy eigenvalues of the potential known as resonances. Similarly, SS is a special discrete energy where $\psi(x)$ becomes outgoing plane wave on both sides of the potential. Thus, at a discrete energy eigenvalue $\psi(x)$ acquires a special asymptotic behaviour. Otherwise, one may call these various kinds of eigenvalues as generalized eigenvalues where eigenfunctions need not necessarily belong to an underlying Hilbert space.

Based on four exactly solvable models (Scarf II, the Dirac delta, the square well and an exponential potential) of CPTSSP and several others solved numerically, we conjecture that, in a potential, the SS is single and it is the upper bound to complex conjugate pairs of eigenvalues ${\cal E}_l \pm i \gamma_l$ $({\cal E}_l \lessapprox  E_*)$ in the potential irrespective of whether the real part of $V(x)$ is a well or a barrier in the parametric regimes of broken PT-symmetry. We propose to  construct a function $F(k)=1/t(k)$, we plot the contours of $\Re [F(k_1,k_2)]=0$ and  $\Im [F(k_1,k_2)]=0$ in the upper  $(k_2>0)$ complex  plane. We collect their points of intersection $(k_1, k_2)$ which help us in finding the complex roots of the type $\pm k_1+ ik_2$ which give rise to three types of  discrete energies ($E=(\pm k_1 + i k_2)^2$: negative real, SS and CCEPs) of a CPTSSP in a convenient way. 

Alternatively, one may also use the  elegant  $2 \times 2$ transfer-matrix ${\cal M}$ method [2,3] of coherent scattering at a complex PT-symmetric medium from left and right. Diagonal elements of ${\cal M}(k)$ and  the two port scattering matrix are written as [13]
\begin{multline}
{\cal M}_{11}(k)=t(k)-r_L(k) r_R(k)/t(k),  \\ \hspace*{-0.3 cm} {\cal M}_{22}(k){=}1/t(k),  S(k){=}t^2(k){-}r_L(k)r_R(k).
\end{multline}
Wherein the zeros of ${\cal M}_{22}(k)=F(k)$ where $\Im(k)>0$ (upper plane) will give us discrete spectra of three types. On the other hand,  physical zeros   of ${\cal M}_{11}(k)$ are in the lower complex plane ($k_2<0$). This is so because in a CPTSSP, the entries of the transfer matrix follow  the property that ${\cal M}^*_{11}= {\cal M}_{22}$ [1,3]. Consequently, a zero of ${\cal M}_{11}$ is always accompanied by the zero of ${\cal M}_{22}$. Further, an SDSS where ${\cal M}_{22}=0$ in a PT-symmetric potential always corresponds to the time-reversed SDSS where ${\cal M}_{11}=0$. While in a generic non-PT-symmetric complex potential a spectral singularity (coherent laser) [7,8,9] and a time-reversed SDSS (CPA) can occur separately, in a PT-symmetric potential they always occur simultaneously, which corresponds to the self-dual spectral singularity (SDSS), i.e., to combined CPA-laser [3] action. SDSS which occur in CCPTSSP implies $T(-k_*)=\infty=T(k_*)$, otherwise an SS in other non-Hermitian potentials implies an SS at $k=k_*$ and $T(-k_*)\ne T(k_*)$; one of these transmission probabilities is infinite but the other one is finite. See interesting examples of these two types of spectral singularities in [15]. In this paper, all spectral singularities (SSs) are actually SDSSs.

In the following, first we present  the complex PT-symmetric version of Scarf II [12] potential whose results are analytic and explicit.  Next, we present three more solvable CPTSSPs whose results are analytic but implicit. Our method of contour plot in complex-$k$ plane $(k_1,k_2)$ works in general. Since negative energy bound states and their evolution as $
V_2$ varies, have been well studied in terms of exceptional points
earlier [16,21] for potentials in Eqs. (16,20), in the following we do this study only for our exponential potential in Eq.  (24) that is new.
\begin{widetext} \noindent
\section{Complex PT-symmetric Scarf II: CCPE and single SDSS}
Complex PT-symmetric Scarf II potential is written as [5]

\begin{equation}
V(x)=V_1 \mbox{sech}^2 x + i V_2 \mbox{sech}x \tanh x, V_1,V_2 \in \Re,
\end{equation}
where $V_1<0$ means the real part is a potential well, otherwise it is a potential barrier. Setting $2 \mu = 1 =\hbar^{2} $, let us define 
\begin{equation}
p=\frac{1}{2}\sqrt{|V_2|-V_1+1/4}, \quad q=\frac{1}{2}\sqrt{|V_2|+V_1-1/4}, \quad s=\frac{1}{2}\sqrt{1/4-V_1-|V_2|}.
\end{equation}
When $V_1<0$ such that $s$ is real, this CCPTSSP is known to have two branches of finite real discrete spectrum given as [5,11]
\begin{equation}
E_{n^\pm}=-[n_+ +1/2-(p \pm s)]^2, \quad n^{\pm}=0,1,2,....[p \pm s -1/2].
\end{equation}
In the above case (6), PT-symmetry is exact (un-broken) and energy eigenstates are also eigenstates of PT such that PT $\psi=\psi$.
When PT-symmetry is broken $q$ becomes real, there is a phase transition of eigenvalues from real to CCPEs as
\begin{equation}
E_n=-[n+\frac{1}{2}-(p \pm iq)]^2, \quad |V_2| > -V_1+1/4=V_{EP}, \quad n=0,1,2,...[p-1/2].
\end{equation}
In this case eigenfunctions flip under PT as PT$\psi_+=\psi_-$.

$V_{EP}$ is called Kato's exceptional point (EP) in the $V_2$-evolution of  real eigenvalues where various pairs of real eigenvalues coalesce to become CCPEs. It has also been shown that when $1/2-p=-m, m=0,1,2,...$ the transmission  $T(E)$ and reflection $R(E)$ probabilities of the Scarf II potential become infinite at a real positive energy given by
\begin{equation}
E_*=\frac{1}{4}[|V_2|-V_1+1/4]>0 \quad \mbox{if} \quad V_2=|V_{*m}|=V_1+ 4m^2+4m+\frac{3}{4},\quad  m=0,1,2,,...
\end{equation}
It can be seen from Eq. (8) that $E_*=k_*^2$ where $k_*=\pm q$. It has been pointed out [13] that the SS $E_*$ is nothing but the last of CCPE in (4) which becomes real and positive, the eigenfunction at this discrete eigenvalue has been shown [13] be plane waves at asymptotic distances on both sides of the potential (4).

If $V_1>0$,  all critical values $V_{*m}$ are possible  for $m=0,1,2,...$ So as $V_2$ increases and crosses them, a new pair of CCPEs is created. When $V_2=V_{*m}$, there is one SS and $m$ number of CCPEs all are calculable from Eq. (7). But when $V_1<0$, $V_{*m}$ exist only after EPs, so there are more than $m$ number of CCPEs along with one SDSS when $V_2=V_{*m}$. The extra (initial) CCPEs arise due to crossing of EPs where some pairs of real discrete eigenvalues coalesce to  make a phase transition (spontaneous breaking of PT-symmetry) to CCPEs.

For fixed value of $V_1>0$, when $V_2$ crosses $V_{*m}$ by becoming $V_2=V_{*m}+\epsilon, \epsilon>0$,the  SDSS disappears and a new pair of CCPE is created so there are $m+1$ pairs of CCPEs. But when $V_2=V_*-\epsilon$, there is no SDSS and there are $m$ CCPEs. This explains the proposed phenomenon of splitting of SDSS [1]. We would like to add that in case we have a real well in $V(x)$ (4) (when $V_1<0$), for SDSS values of $V_{*m}>V_{EPl}$. 

All the expressions  for Scarf II are explicit and most simple to see the acclaimed results, however we present the Table 1 which displays the scenario for $V_1=0, 5$ and $-5$ in a more convenient way, when $V_2$ is varied and takes four consecutive special critical values $V_{*m}$. Row nos. $\{5,10,17\}$ display the splitting of the  SDSS $E_*$ in to CCPEs when $V_2=V_*+0.1$. Also notice that a  CPTSSP has at most one SDSS and it doesn't exist when the potential has real discrete spectrum  (see row no 11, when PT-symmetry is un-broken: $V_2<-V_1+1/4)$. Also notice that when $V_1<0$, SDSS does not occur alone even for the first critical $V_*$, it is  accompanied by CCPEs which are due phase transition of real discrete eigenvalues to CCPEs (see Row no. 13). In the whole of the Table I, notice that SDSS  is the upper bound to the real part of CCPEs.

	\begin{table}[]
		\centering
		\caption{The evolution of discrete eigenvalues for Scarf II (4), when $V_1$ is fixed and $V_2$ admits first four  critical values $V_{*m}, m=0,1,2,3,...$ (8). If $V_2=V_{*m}$ there is one SDSS and $m$ or more number of CCPEs. For splitting of the SDSS see row nos. $\{5,10,17 \}$. When real part of $V(x)$ is a well ($V_1<0$)
			SDSS does not occur alone, it occurs with at least one CCPE. Also notice that a real  bound state eigenvalue and SDSS are mutually exclusive. In all these cases 
			$E_*$ is upper bound to the real part of CCPEs. We have set $2\mu/\hbar^2=1 (eV \AA^2)^{-1}$, which corresponds to $\mu \approx 4 m_e$ ($m_e$ is mass of electron) where $\mu$, $E_n$, $E_*$, $V_1$ and $V_2$ are in $eV$.
			 \\}
		\label{my-label}
		\begin{tabular}{|c|c|c|c|c|}
			\hline
			S. No.&	$V_1$ & $V_2{=}V_*,V{_*+}0.1$   & $E_*$ & $E_n$ 
			\\ \hline
			1&	0 & 0.75 & 0.125 & -- \\ \hline
			2& 0 & 8.75 & 2.125 & $ 1.125 \pm i 2.915 $ \\ \hline
			3& 0 & 24.25 & 6.125 & $5.125 \pm i 4.949 , 2.125 \pm i 9.892 $ \\ \hline
			4& 0 & 48.75 & 12.125 & $11.125 \pm i 6.964, 8.125 \pm i 13.982, 3.125  \pm i 20.892 $ \\ \hline
			5& 0& 48.85 & - & $12.150 \pm i 0.024, 11.142 \pm i 6.996, 8.135 \pm i 13.967,3.128 \pm i 20.939$ \\ \hline
			6& 5 & 5.75 & 2.625 &- \\ \hline
			7& 5 & 13.75 & 4.625 & $3.625 \pm i 4.301 $ \\ \hline
			8& 5 & 29.75 & 8.625 & $7.625\pm  i 5.873 , 4.625\pm i 11.747 $ \\ \hline
			9& 5 & 53.75 & 14.625 & $ 13.625\pm i7.648, 10.625\pm i 15.297, 5.625\pm i 22.945$ \\ \hline
			10& 5 & 53.85 & -& $ 14.650 \pm i 0.027 , 13.642\pm i7.682, 10.635 \pm i 15.337, 5.628\pm i 22.992$ \\ \hline
			11& -5 & 5.24 & - & -1.367,~~ -1.143,~~ -0.028,~~ -0.004 \\ \hline
			12& -5 & 5.50 & -& $-1.235\pm i 0.569, 0.043\pm i 0.069$ \\ \hline
			13& -5 & 19.75 & 3.625 & $2.625 \pm i 3.002, -0.375 \pm i 7.615$ \\ \hline
			14& -5 & 43.75 & 9.625 & $8.625 \pm i 6.204, 5.625 \pm i 12.409 , 0.625 \pm i 18.614$ \\ \hline
			15& -5 & 75.75 & 17.625 & $16.625 \pm i 8.396 , 13.625 \pm i 16.792, 8.625 \pm i 25.189, 1.625 \pm i 33.585$ \\ \hline
			16& -5 & 115.75 & 27.625 & \specialcell{$2.625 \pm i 52.559, 11.625 \pm i 42.047, 18.625 \pm i 31.535, 23.625 \pm i 21.023$ \\ $26.625 \pm i 10.511$} \\ \hline
			17& -5 & 115.85 & - & \specialcell{$27.65 \pm i 0.023, 26.645 \pm i 10.540, 23.640 \pm i 21.057, 18.636 \pm i 31.573$, \\ $11.636 \pm i 42.090, 2.627 \pm i 52.607$}  \\ \hline
			
		\end{tabular}
	\end{table}

\begin{figure}
	\includegraphics[width=7. cm,height=7. cm]{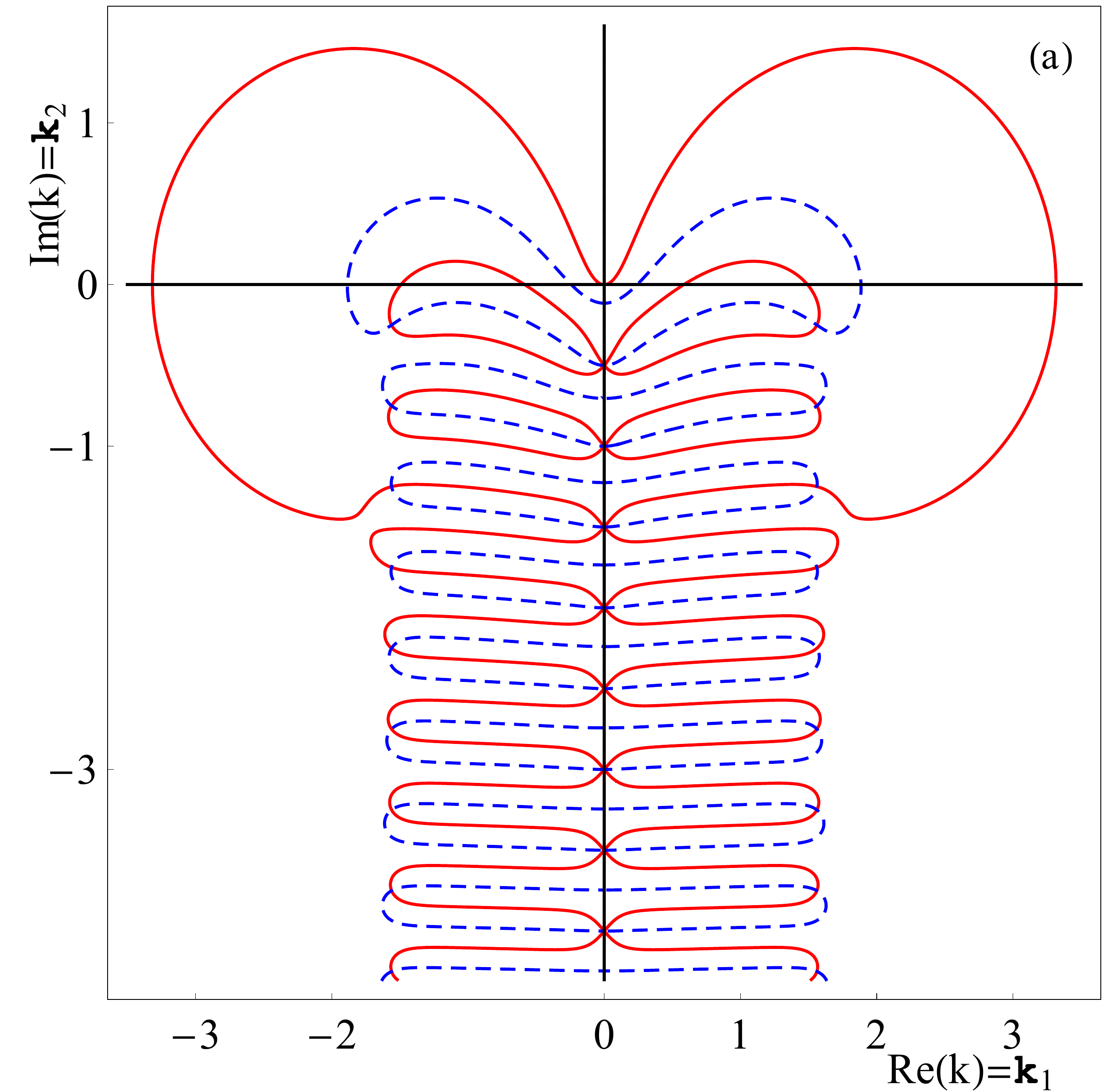}
	\hspace{.5 cm}
	\includegraphics[width=7. cm,height=7. cm]{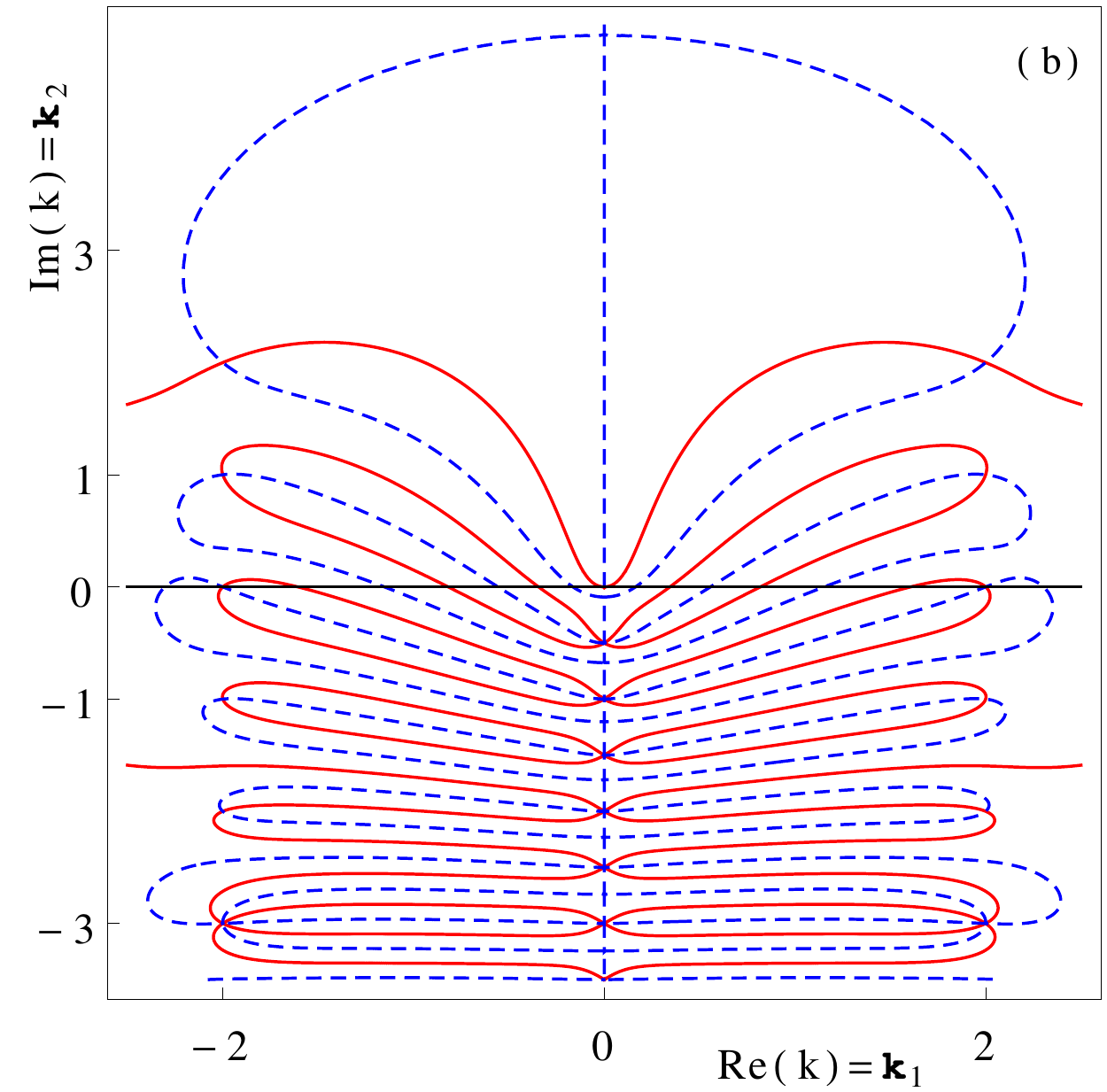}
	\caption{ Contour plots of real and imaginary parts of $F(k)={\cal M}_{22}(k_1+ik_2)=0$ for Scarf II (4) in complex $k$-plane. (a) when $V_1=5$ and $V_2=5$  and $(b): V_1=-5, V_2=19.75$ (see row no. 13 in Table I).  Solid red lines and  dashed  blue lines  depict real and imaginary parts of $F(k)$, respectively. In (a),  notice that solid  and dashed curves are intersecting  only for $k_2<0$, so no physical zero giving rise to discrete eigenvalues (SDSS or CCPE). In (b), these pairs of curves are intersecting three times for $k_2\ge 0$. For the first point of intersection $k_2=0$ (SDSS), two upper ones denote two CCPEs. Units of physical quantities are as in Table I. Unit of $k$ is $\AA^{-1}$.}
\end{figure}
Beautiful expressions of reflection $r(k)$ and transmission $t(k)$ amplitudes for Scarf II (4) are available in Ref. [12] in terms of their parameters $a$ and $b$. Using $a=(p+q-1)/2$ and $b=i(p-q)/2$ in them, we investigate the poles of $t(k)$ or zeros of $F(k)={\cal M}_{22}(k)=1/t(k)$ in complex $k$-plane.
Fig. 1, depicts the extraction of CCPE and SDSS from the intersection of contour plots of $\Re[F(k_1,k_2)]=0= \Im [F(k_1,k_2)]$ in complex $k$-plane. In Fig. 1(a), there is no intersection point lying in upper half plane, so there are no CCPEs. This is the case which is devoid of any discrete spectrum.  Fig 1(b) presents the case when there are physical $k$-poles giving one SDSS and  two CCPEs (see row no. 13 in Table I).

By invisibility [14] of a potential for an energy $E=E_i$, $T(E_i)=1$ and one or both of $R_L(E_i)$ and $R_R(E_i)$ vanish. In case only one of the reflectivities vanishes it is called uni-directional invisibility. We find that Scarf II (4) becomes unidirectionally invisible for at most  one special energy.

\section{Complex PT-symmetric Dirac delta potential} 

\begin{table}[]
	\centering
	\caption{The same as in Table I for the complex PT-symmetric Dirac delta potential (9). Here, $a=1 \AA$. \\} 
	\label{my-label}
	\begin{tabular}{|c|c|c|c|c |}
		\hline
		S. No. &	$V_1$ & $V_2{=}V_*,V{_*+}0.1$   & $E_*$ & $E_n$
		\\ \hline
		1&	0 & 1.110 & 0.616 & - \\ \hline
		2&	0& 3.332 & 5.552 & $1.921 \pm i 0.663$  \\ \hline
		3&	0& 5.553 & 15.421 & $7.550  \pm i 1.920$, $2.415 \pm i 0.291$ \\ \hline 
		4&	0& 7.775 & 30.226 & $ 17.222 \pm i 2.965, 9.424 \pm i 1.455,  2.457 \pm i  0.139 $ \\ \hline
		5&	0& 7.875 & -- & $ 30.251 \pm i 0.137, 17.365 \pm i  3.063, 9.460 \pm i  1.412, 2.457 \pm i  0.135 $ \\ \hline
		6&	5 & 5.394 & 2.048 & - \\ \hline
		7&	5 & 6.449 & 8.291 & $2.116 \pm i 0.016$  \\ \hline
		8&	5 & 7.932 & 18.958 & $8.606 \pm i 0.136, 2.194 \pm i 0.0274$  \\ \hline
		9&	5 & 9.668 & 34.238 & $19.591 \pm i 0.397, 8.908 \pm i 0.207, 2.261 \pm i 0.031$  \\ \hline
		10&	5 & 9.768 & -- & $34.284 \pm i  0.048, 19.627 \pm i  0.413, 8.923 \pm i  0.208, 2.264 \pm i  0.031   $  \\ \hline
		11&	-5 & 0 & -- & -6.163,~~ -6.332 \\ \hline
		12 & -5 & 2 & - & $5.250 \pm i 5.000$ \\ \hline
		13&	-5 & 5.571 & 3.020 & $1.509 \pm i 13.929$  \\ \hline
		14&	-5 & 6.979 & 11.855& $ 5.928 \pm i 17.447, 2.874 \pm i 0.037 $  \\ \hline
		15&	-5 & 8.788 & 26.118 & $ 11.273 \pm i 0.275, 13.058 \pm i 21.971, 2.748 \pm i  0.046 $  \\ \hline
		16&	-5 & 10.776 & 45.560 & $ 25.085 \pm i  0.744, 22.781 \pm i 26.939, 10.811 \pm i  0.316, 2.663 \pm i  0.041 $  \\ \hline
		17&	-5 & 10.876 & -- &\specialcell{ $ 45.510 \pm i 0.075, 25.032 \pm i 0.764, 23.323 \pm i 27.190,
		10.793 \pm i  0.315,$ \\ $ 2.659 \pm i 0.041 $}  \\ \hline
	\end{tabular}
\end{table}

This potential is expressed as
\begin{equation}
V(x)=(V_1-iV_2) \delta(x+a) + (V_1+iV_2) \delta(x-a)
\end{equation}
and this has been used to study various features of complex PT-symmetry [16-19]. The reflection  amplitude can be written [16] as 
\begin{equation}
r(k,V_2)=\frac{-ie^{-2ika}[(2kV_2+V_1^2+V_2^2)\sin 2ka +2kV_1 \cos 2ka ]}{2k^2 \cos 2ka+2kV_1 \sin 2ka +i[2kV_1 \cos 2ka +(V_1^2+V_2^2-2k^2)\sin 2ka]}.
\end{equation}
For the incidence from left $r_L(k)=r(k,V_2)$  and from right it is $r_R(k)=r(k,-V_2)$. This demonstrates non-reciprocity of reflection namely, $R_L(k)\ne R_R(k)$. The transmission amplitude [14] is
\begin{equation}
t(k,V_2)=\frac{2k^2e^{-2ika}}{2k^2 \cos 2ka+2kV_1 \sin 2ka +i[2kV_1 \cos 2ka +(V_1^2+V_2^2-2k^2)\sin 2ka]}
\end{equation}
Notice that $t(k)$ is invariant if we change $V_2$ to $-V_2$ (reflection of the potential), so the transmission is reciprocal. One can check that $ R_L(-k)=R_R(k)$ and $T(-k)=T(k).$
 In this potential the  SDSS is single and it has been found [19]  at $k=k_*=\frac{\pi(2m+1)}{4a}$ when $V_1 = 0$ and $V_2=\frac{\pi (2m+1)}{2 \sqrt{2} a}=V_{*m}$, $m = 0, 1, 2, 3..$. If we set $2k^2=V_2^2-V_1^2$ in the denominators of $r(k,V_2)$ and $t(k,V_2)$, the denominator has $(k \cos 2ka + V_1 \sin 2ka)$ as  a factor. Thus, the SDSS is found at,
 
\begin{multline}
 E = E_*=k_*^2 = [V_2^2 - V_1^2]/2, \mbox{where} \quad 2V_1+\sqrt{2[V_2^2-V_1^2]}~\cot[a\sqrt{2[V_2^2-V_1^2}]=0.
\end{multline} 
Further, we numerically check that $|\det(S(k))|=|t^2(k)-r_L(k)r_R(k)|=1$ except when $k=k_*$, at this value $|\det(S(k_*)|$ becomes indeterminate $0/0$ but its limit as $k\rightarrow k_*$ becomes 1. This action of the optical medium with balanced gain and loss is called CPA with lasing [3,7]. 
\begin{figure}
	\includegraphics[width=7. cm,height=7. cm]{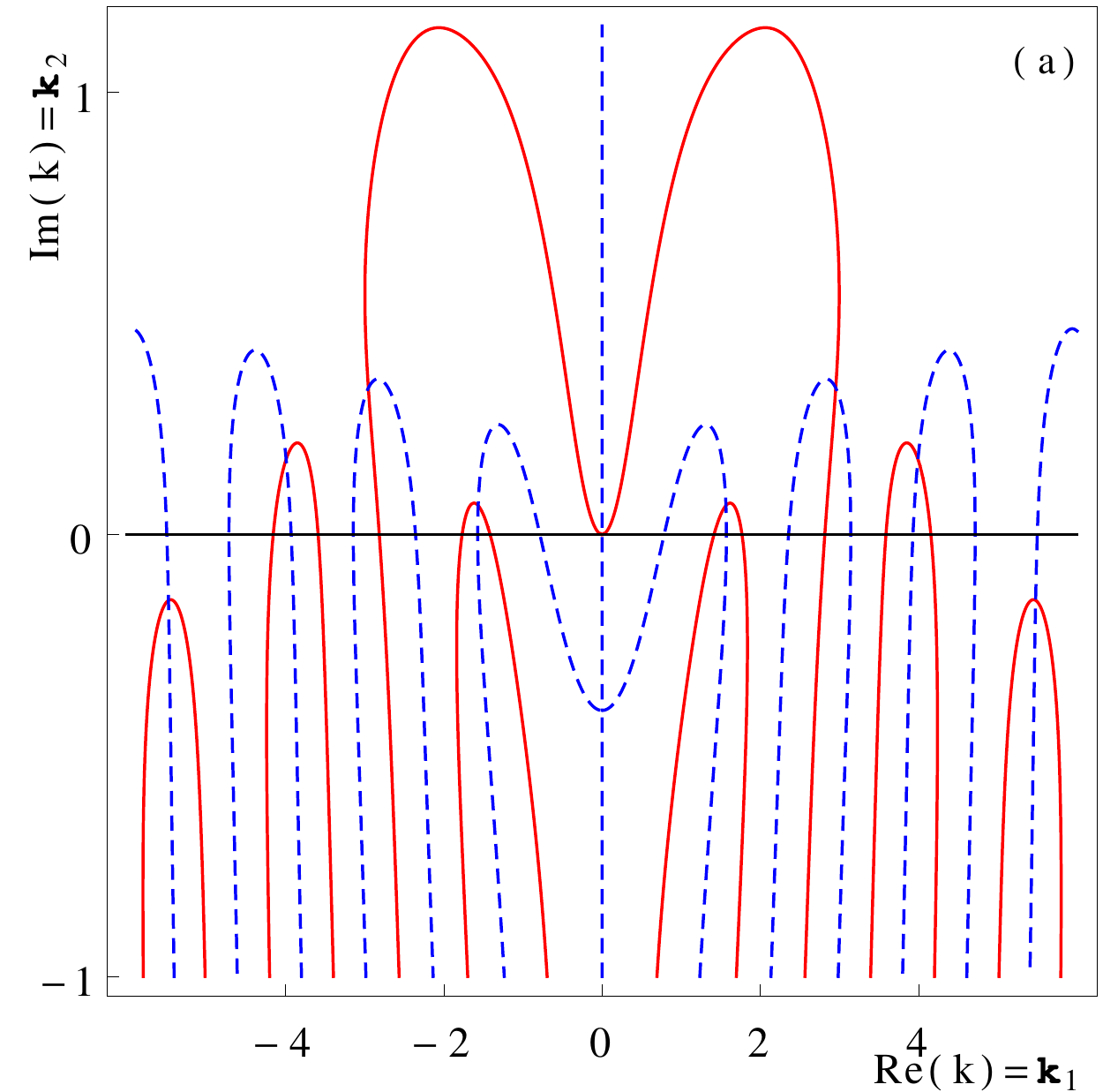}
	\hspace{.5 cm}
	\includegraphics[width=7. cm,height=7. cm]{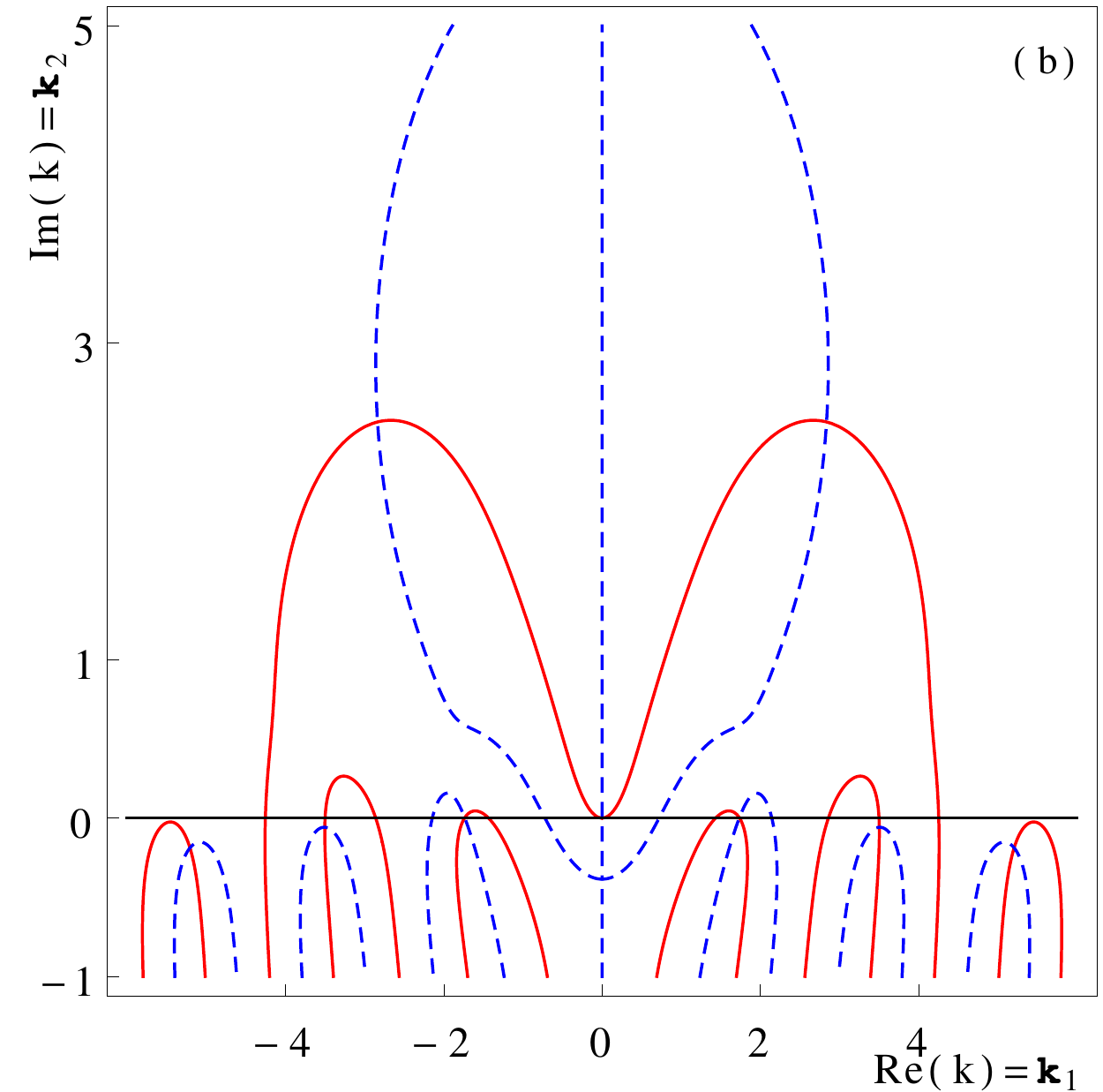}
	\caption{ The same as in Fig. 1 for Dirac delta potential (9) $(a=1 \AA)$. (a) when $V_1=0$, $V_2=5.653$, there are three pairs of roots with $k_2>0$, these are $15.456 \pm i 0.133, 7.663\pm i 1.949, 2.420\pm i 0.280$, The first pair is due to splitting of SDSS that occurs at $E_*=15.421$  when $V_2=5.553=V_*$ (see row no. 3, in Table II). (b) When $V_1=-5$, $V_2= 5.671=V_*+0.1$ there are two  CCPEs: $3.009 \pm i 0.004, 1.789 \pm i 14.177$.  The first one results from  the splitting of SDSS  (see row no. 13 in Table II).}
\end{figure}
Notice that features of the delta potential (9) presented in Table II are the same as that of Scarf II (4). Fig. 2, displays two examples of splitting of SDSS when $V_2=V_*+0.1$.

Not shown here, we also find that the delta potential (9) displays both uni-directional and bi-directional  invisibility, frequently.  When $V_1=0$, bi-directional invisibility exists at $E=E_{in}=\frac{n^2 \pi^2 \hbar^2}{8\mu a^2}$, incidentally these are the discrete eigenvalues of infinitely deep well of width $2a$. When $V_1 \ne 0$, this potential (9) becomes easily unidirectionally invisible at several discrete energies.
 
\section{Complex PT-symmetric square well potential:}

	\begin{table}[]
		\centering
		\caption{The same as in Table I for the complex PT-symmetric square well potential  (13). Here, $a= 2\AA$. \\}
		\label{my-label}
		\begin{tabular}{|c|c|c|c|c|}
			\hline
			S. No.&	$V_1$ & $V_2{=}V_*,V{_*+}0.1$   & $E_*$ & $E_n$ 
			\\ \hline
			1&	0 & 0.519 & 0.284 & -- \\ \hline
			2&	0 & 3.330 & 4.674 & $1.1423 \pm i 2.668$ \\ \hline
			3&	0&  6.946  & 14.172 & $5.950 \pm i4.244, 1.470 \pm i 6.352$ \\ \hline
			4&	0&  11.028  & 28.701 & $15.390 \pm i 5.209, 6.642\pm i 8.698, 1.647 \pm i 10.464 $ \\ \hline
			5&	0&  11.128  & -- & $28.728\pm i 0.139, 15.413 \pm i 5.330, 6.654\pm i 8.806, 1.650 \pm i 10.595 $ \\ \hline
			6&	5& 0.915 & 5.851 & --\\ \hline 
			7&	5& 3.685 & 10.534 & $6.408 \pm i 2.844$ 	\\ \hline 
			8&	5& 7.259 & 20.368 & $11.520 \pm i 4.245, 6.593 \pm i 6.556$ 	\\ \hline
			
			9&	5& 11.289 & $34.845$  & $21.124 \pm i 5.162, 11.952 \pm i 8.731, 6.714 \pm i10.684$  \\ \hline
			10&	5& 11.389 & $-$  & $34.864 \pm i0.136,21.139\pm i 5.284, 11.967\pm i 8.839, 6.717\pm i 10.786$  \\ \hline
			11 & -5 & 2.000 & - & -0.083,~~ -0.918,~~  $-3.823 \pm i 1.601$ \\ \hline
			12 & -5 & 5.900 & - &  $0.621 \pm i  3.899 $,  $-3.560 \pm i 5.466 $ \\ \hline
			13&	-5 & 6.000 & 8.033&  $0.646 \pm i3.998, -3.556 \pm i5.566$  \\ \hline
			14&	-5& 10.383 & 22.578 & $9.732 \pm i 5.195, 1.445 \pm  i 8.429, -3.381 \pm i9.941$  \\ \hline
			15&	-5& 14.960 & 42.137 & $23.995\pm i 5.826, 10.808\pm i 10.308, 1.947 \pm i 13.119, -3.263\pm i 14.533 $
			\\ \hline
			16& -5 & 19.738 & 66.662 & \specialcell{$ 43.334\pm i 6.246, 25.054\pm i 11.521, 11.551 \pm i 15.491, 2.294\pm i 18.010,$\\ $-3.179\pm i 19.329 $} \\ \hline
			17& -5 & 19.838 & -- & \specialcell{$ 66.685\pm i 0.137, 43.357\pm i 6.372, 25.072\pm i 11.638, 11.563 \pm i 15.598,$\\ $ 2.299\pm i 18.112, -3.177\pm i 19.430 $} \\ \hline
		\end{tabular}
	\end{table}

Piece-wise constant complex PT-symmetric square well potential which is written as 
\begin{equation}
V(|x|<a)=[V_1+i V_2 ~ \mbox{sgn}(x)], V(|x|>a)=0.
\end{equation}
has been first discussed [20] for demonstrating the non-reciprocity [6] of reflection from left and right. Let us define $p,q= \sqrt{\frac{2\mu}{\hbar^2}(E-V_1 \pm i V_2)}$, $k=\sqrt{\frac{2\mu E}{\hbar^2}}$. For incidence from left, the solution of one dimensional time-independent Schr{\"o}dinger equation (2) can be written as
\begin{eqnarray}
\psi(x\le -a)&=& A e^{ikx} +B e^{-ikx}, \psi(-a<x\le 0) =C \sin px +D \cos p x, \\ \nonumber
\psi(0<x\le a)&=& (p/q) C \sin qx+ D \cos qx, \psi(x>a)= F e^{ikx}. 
\end{eqnarray}
Matching the scattering solution $\psi(x)$ and its derivative at $x=0$, we find equations for $A,B,C,D$ and $F$. From these equations, we obtain the reflection amplitude $r(k,p,q)=$
\begin{equation}
\frac{B}{A}=\frac{[ q(k^2-p^2) \sin P \cos Q +p (k^2-q^2) \cos P \sin Q+ik(p^2-q^2) \sin P \sin Q] e^{-2ika}}{2ikpq \cos P \cos Q +p(k^2+q^2)  \cos P \sin Q + q(p^2+k^2) \sin P \cos Q-ik(p^2+q^2) \sin P \sin Q}
\end{equation}
and transmission amplitude, $t(k,p,q)=$
\begin{equation}
\frac{C}{A}{=}\frac{2ikpq e^{{-}2ika}}{2ikpq \cos P \cos Q {+}p(k^2{+}q^2)  \cos P \sin Q{+} q(p^2{+}k^2)  \sin P \cos Q{-}ik(p^2{+}q^2) \sin P \sin Q},
\end{equation}
where $P=pa$ and $Q=qa$.
\begin{figure}
	\includegraphics[width=7. cm,height=7. cm]{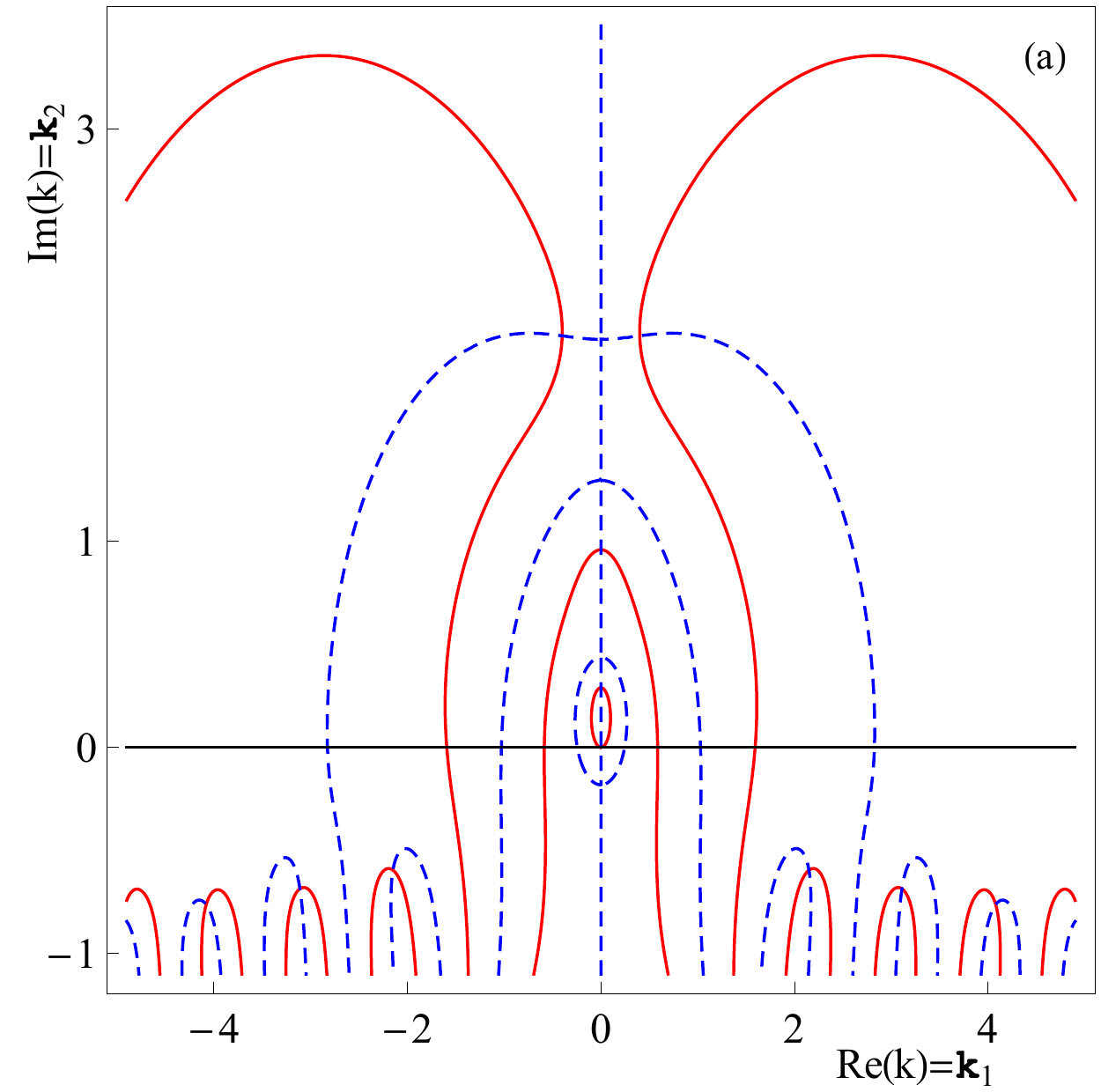}
	\hspace{.5 cm}
	\includegraphics[width=7. cm,height=7. cm]{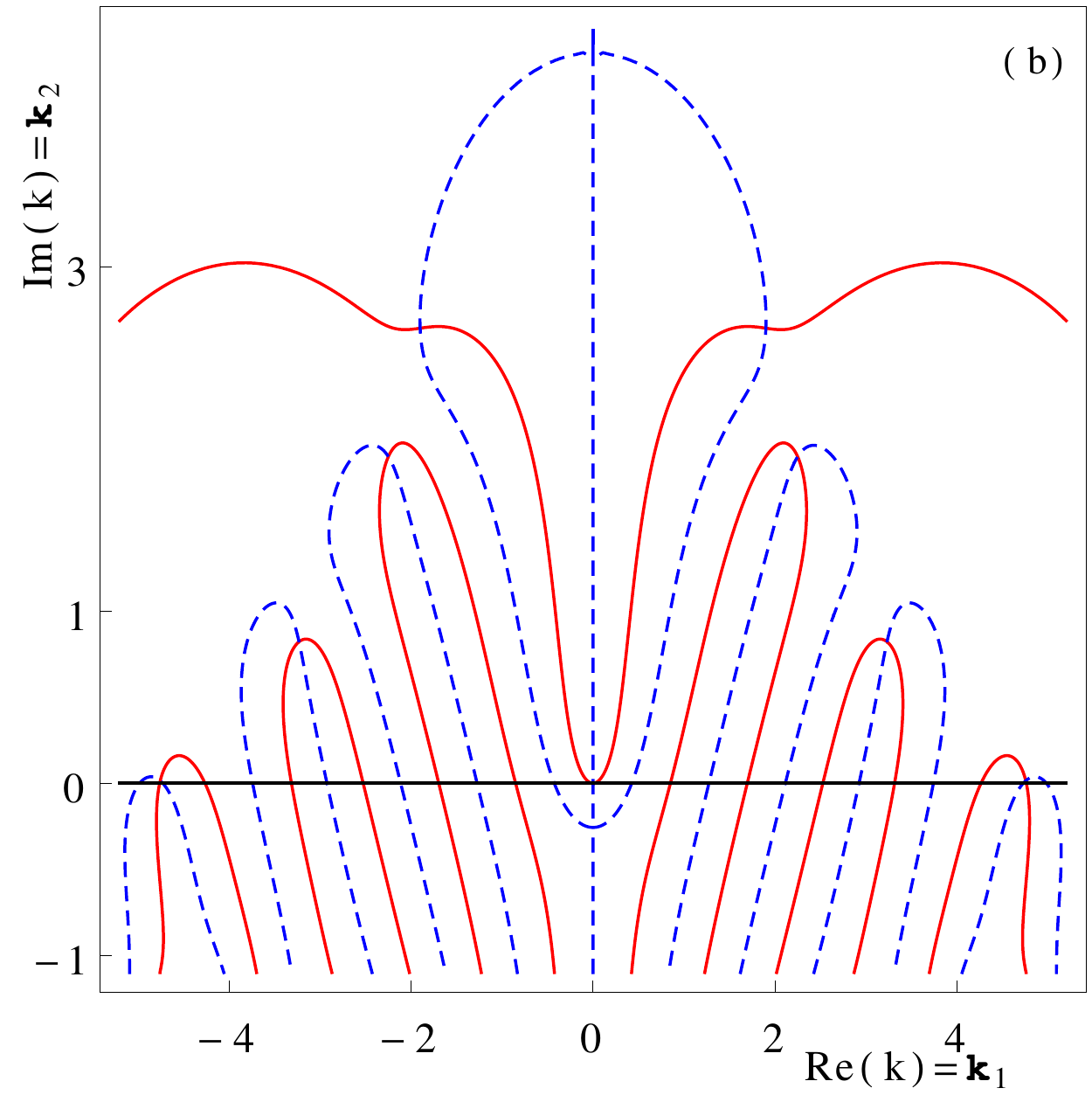}
	\caption{The same as in Fig. 1 for square well potential (13) ($a=2 \AA$). (a) when $V_1= -5$ and $V_2=2.0$, there are two negative real discrete eigenvalues  as the dashed (blue) and solid (red) curves intersect on upper $y-$axis ($k_1=0, k_2 >0$) and we have one CCEP (see row no. 11, in Table III). (b) When $V_1=-5$, $V_2=10.483$ there are four pairs of roots with $k_2>0$, these pairs are $22.6139 \pm i 0.134, 9.762+5.309i, 1.45\pm 8.530i, -3.377 \pm i 10.041$. The first pair results from the splitting of SDSS that occurs at $E_*=22.578$  when $V_2=10.383=V_*$ (see row nos. 14, in Table III).}
\end{figure}
The table III, presents the same scenario of Table I and II for the square well potential (13). Here unlike potentials (4,9), there could be mixture of negative real discrete eigenvalues and CCPE (see row no. 11 in Table III). In Fig 3, two cases of splitting of SDSS are presented. We also observe uni-directional invisibility at several energies in model.

 		\begin{table}[]
 		\centering
 		\caption{ The same as in  Table I for the complex PT-symmetric exponential well  ($a=2 \AA $) (17). Notice the $\#$ sign in row nos. $\{9,14,16\}$ where in  over all 61 cases presented in 4 Tables, only here, ${\cal E}_l$ is slightly  greater than $E_*$, rendering SDSS as rough upper bound to ${\cal E}_l$ $({\cal E}_l \approx  E_*)$. \\}
 		\label{my-label}
 		\begin{tabular}{|c|c|c|c|c|}
 			\hline
 			S. No. &$V_1$ & $V_2$   & $E_*$ & $E_n$ 
 			\\ \hline
 			1& 0 & 1.330 & 0.225 &- \\ \hline 
 			2&	0 & 14.245& 3.400   & $3.180 \pm i 5.606 $  \\ \hline
 		    3& 0 & 40.250 &9.943 & $10.733 \pm i 8.610, 7.760 \pm i 21.637$ \\ \hline
 		    4& 0 & 79.253 & 19.816& $21.581 \pm i 11.464, 20.199 \pm i 27.026, 13.442 \pm i 48.951$
 		    \\ \hline
 		    5& 0 & 79.353 & -& $19.487 \pm i 0.023, 31.609 \pm i 11.681, 20.221 \pm i 27.067, 13.455 \pm i 49.024 $ \\ \hline
 		    	6&	5 & 5.534 & 4.129 & -  \\ \hline
 		    	7&  5 & 20.470 & 6.997 & $7.634 \pm i 8.305$ \\ \hline
 		    	8& 5 & 46.232 &  13.467 & $14.028 \pm i 10.212, 12.392 \pm i 25.488$ \\ \hline 
	9& 5 &  $ 86.139$ & $^{\#} 23.3298$ & $ ^{\#}25.502 \pm i 12.769, 24.517 \pm i  29.547, 18.154 \pm i 53.401 $ \\ \hline 
	10& 5 & 86.239 & -- & $23.362 \pm i 0.026, 25.531 \pm i 12.807,  24.540 \pm i 29.601 , 18.168 \pm i 53.474$ \\ \hline 
 		    
 			11& -5 & 3 & - & -1.2334,~~ -1.554  \\ \hline
 			12& -5 & 3.2& -& $-1.390 \pm i 0.3611$ \\ \hline
 			13& -5 & 3.381 & 0.035 & $-1.370 \pm i 0.537  $ \\ \hline
 			14&	-5 & 32.473 & $^{\#}6.441$ &  $^{\#}6.651 \pm i 6.796, 3.113 \pm i 17.229$  \\ \hline
 		15& -5 & 71.865 &  16.326 & $ 17.666 \pm i 10.449 , 15.875 \pm i 24.310, 8.720 \pm i 44.185 $ \\ \hline
	16& -5 & 124.100 &  $ ^{\#} 29.569 $ & \specialcell{ $^{\#} 31.941 \pm i 13.782, 31.437 \pm i 30.841, 26.837 \pm i 52.641,$ \\ $ 15.213 \pm i  82.867  $ }\\ \hline 	
		17& -5 & 124.200 &  -- &\specialcell{ $29.601 \pm i  0.022, 31.970 \pm i  13.814 , 31.462 \pm i  30.885, 26.856 \pm i  52.698,$ \\$ 15.224 \pm i  82.943 $} \\ \hline

 			18&	-60 & 10 & - & -43.25, -30.82, -20.25, -14.95, -9.19, -6.29, -3.16, -1.75, -0.44, -.065  \\ \hline
 			19&	-60& 14 &- & \specialcell{ -39.32, -33.70 ,$-17.64 \pm i1.03, -7.73 \pm i0.91, -2.43 \pm i0.52,$\\$ -0.20 \pm i0.15$} \\ \hline
 			20& -60 & 83 & 8.551 & \specialcell{ $8.51 \pm i 6.75, 5.87 \pm i 14.99, -0.20 \pm i 25.17, -11.235 \pm i 38.20,$ \\$ -30.84 \pm i 56.27$ }\\ \hline
 		\end{tabular}
 \end{table}
 \section{ Complex PT-symmetric exponential scattering potential}
 	
 This is a new CPTSSP to be expressed as
 \begin{equation}
 V(x)= [V_1 +iV_2~ \mbox{sgn}(x)] ~  e^{-2{|x|/a}}.
 \end{equation}	
 {\bf Scattering states:}\\
 Let us introduce $p,q=a\sqrt{\frac{2\mu}{\hbar^2}(-V_1 \pm i V_2)}$ and $s=ka$, $k=\sqrt{\frac{2 \mu E}{\hbar^2}}$, for solving the one-dimensional time-independent Schr{\"o}dinger equation (2) with this exponential potential (17). Using the solvability of the Schr{\"o}dinger equation for this potential in terms of cylindrical Bessel functions  $J_{\pm ika}(q e^{|x|/a})$, we can write the scattering eigenstates for the incidence from left as
 \begin{equation}
 \psi(x<0) = A\alpha (p/2)^{-is}  J_{is} (p e^{x/a}) + B \alpha^* (p/2)^{is} J_{-is} (p e^{x/a}), \psi(x \ge 0) = C \alpha^* (q/2)^{is} J_{-is}(q e^{-x/a}).
\end{equation}
Here $\alpha=\Gamma(1+is)$.
Owing to the property that $J_\nu(z) \approx (z/2)^{\nu}/\Gamma(1+\nu)$, when $z$ is very small, we see that $\psi(x<0)$ behaves as combination of incident and reflected waves $A e^{ikx} + B e^{-ikx}$ when $x \sim -\infty$ and $\psi(x \ge 0)$ when $x \sim \infty$ behaves as transmitted wave travelling from left to right. By matching $\psi(x)$ and its derivative at $x=0$, we find $B/A$ and $C/A$ which define the reflection amplitude $r(k,p,q)$ 

\begin{equation}
r(k,p,q)= -(p/2)^{-2is}\frac{\Gamma(1+is)}{\Gamma(1-is)} ~\left ( \frac{q J_{is}(p) J'_{-is}(q)+ p J_{-is}(q) J'_{is}(p)}{q J_{-is}(p) J'_{-is}(q) + p J_{-is}(q) J'_{-is}(p)} \right)
\end{equation}
and transmission $t(k,p,q)$ amplitude as
\begin{equation}
t(k,p,q)= (pq/4)^{-is}\frac{\Gamma(1+is)}{\Gamma(1-is)} ~\left ( \frac{2i  \pi^{-1}\sinh \pi s }{q J_{-is}(p) J'_{-is}(q) + p J_{-is}(q) J'_{-is}(p)} \right).
\end{equation}
 
Notice that $r(k,p,q)$ and $r(k,q,p)$ are unequal, the former denotes $r_L(k)$ while the latter is $r_R(k)$. Next, $t(k,p,q)$ being symmetric in $p$ and $q$ ensure reciprocity of transmission and 
so $t_L(k)=t(k,p,q)=t_R(k)$ . \\ 
 \begin{figure}
 	\includegraphics[width=7. cm,height=7. cm]{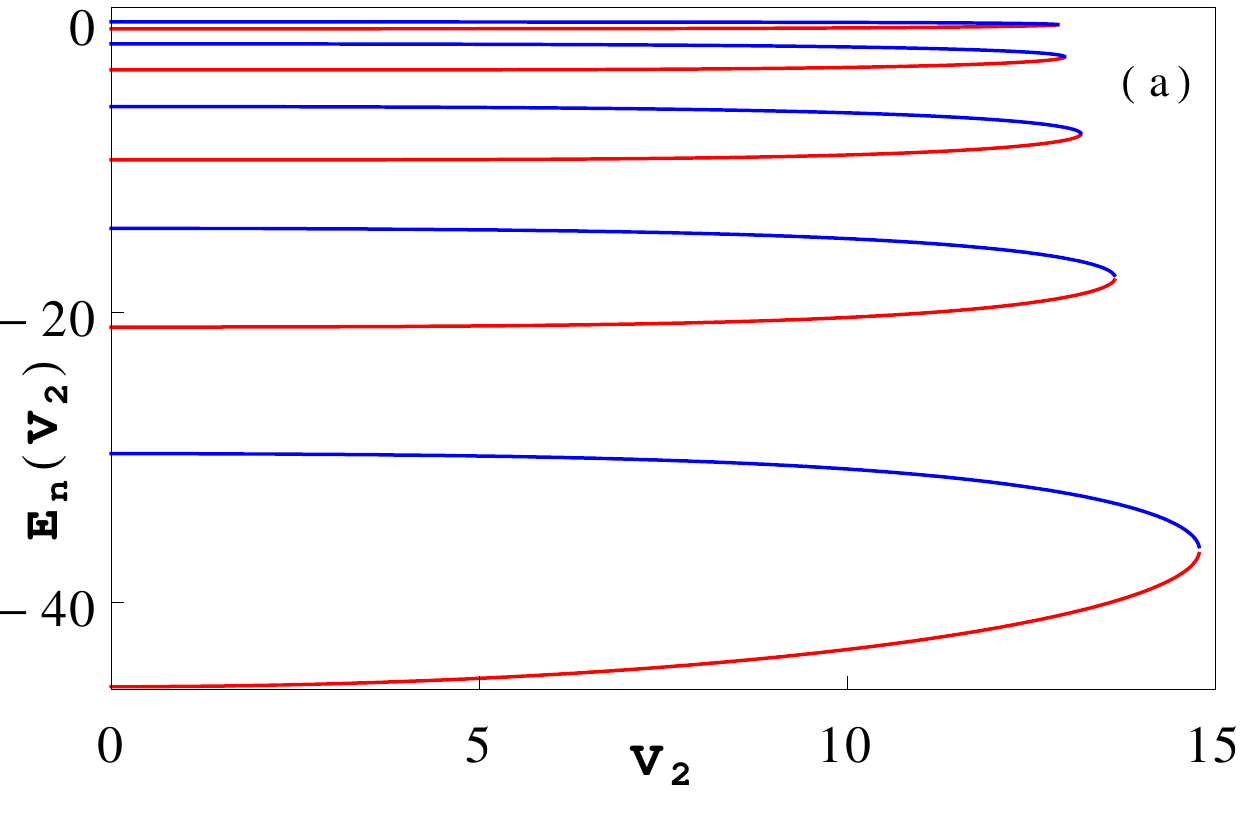}
 	\includegraphics[width=7. cm,height=7. cm]{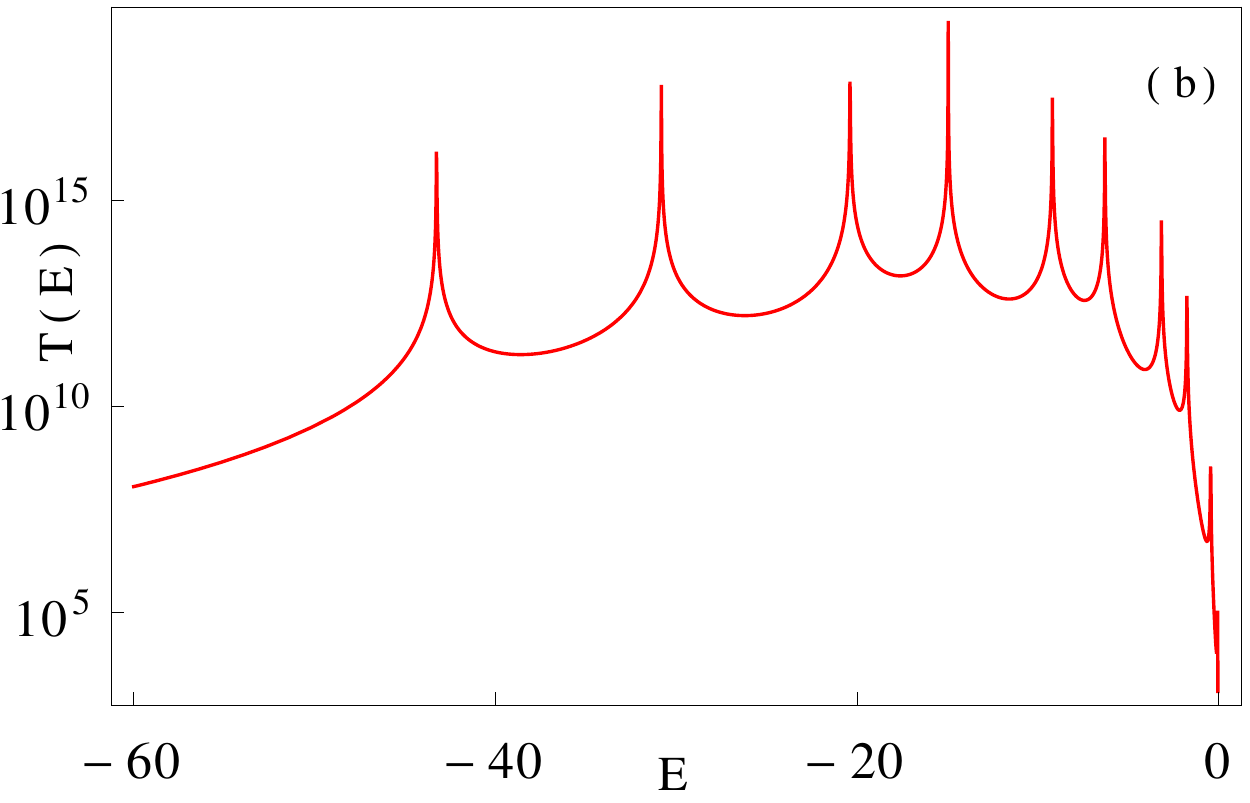}
 	\caption{The evolution of  negative real discrete eigenvalues of the exponential potential (17) $E_n(V_2)$ when $V_1{=}{-}60 eV$ and $a{=}2 \AA$.\hspace*{.5cm} Co-ordinates of the end points are $(12.82, -0.22), ~ ~(12.96,-2.44), ~(13.17,-7.73),$ $(13.63, -17.64), (14.78, -36.40)$ which are in units of $eV$. So EPs of the potential are $V_{EPn} (eV): 12.82, 12.96, 13.17, 13.36$ and 14.78. This structure is temple-like (length of the bottom loop is largest), for a cone-like structure of evolution see [21]. The part (b) presents 10 real discrete negative  eigenvalues which are negative energy poles of $T(E)$ (20) for the case of the potential (see row no. 19 of Table IV).
 	}
 \end{figure}\\
 \noindent  \noindent  
{\bf Real discrete energy bound states:}\\ 
Let us define $\kappa_n =\sqrt{2\mu(-E_n)/\hbar^2}$, out of two linearly independent solutions $J_{\pm \kappa_n a}(p e^{-|x|/a})$  of the Schr{\"o}dinger equation (2), the appropriate solutions vanishing asymptotically are $\psi(x<0)= A J_{-i\kappa_n a}(p e^{-x/a})) $ and $\psi(x \ge 0)= B J_{i\kappa_n a}(p e^{x/a}))$. Matching these solutions at $x=0$,
 \noindent
we get energy eigenvalue equation for the real discrete energy eigenvalues of the exponential well as
\begin{equation}
p J_{\kappa_n a}(q) J'_{\kappa_n a} (p)+ q J_{\kappa_n a}(p) J'_{\kappa_n a} (q)=0,
\end{equation}
which are nothing but negative energy (physical) poles of $r$ and $t$ in Eqs. (19,20).
\begin{figure}
	\includegraphics[width=7. cm,height=7. cm]{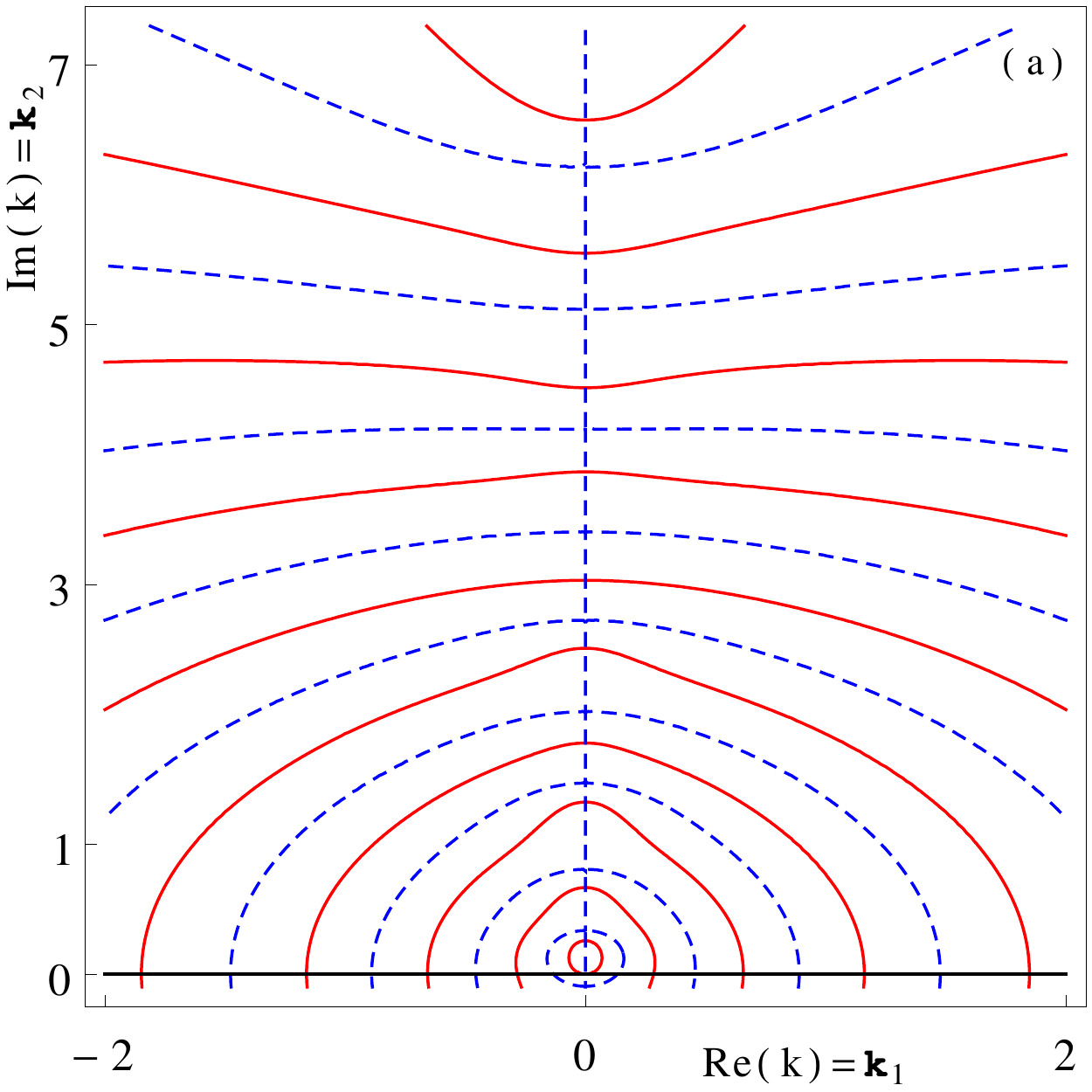}
	\hspace{.5 cm}
	\includegraphics[width=7. cm,height=7. cm]{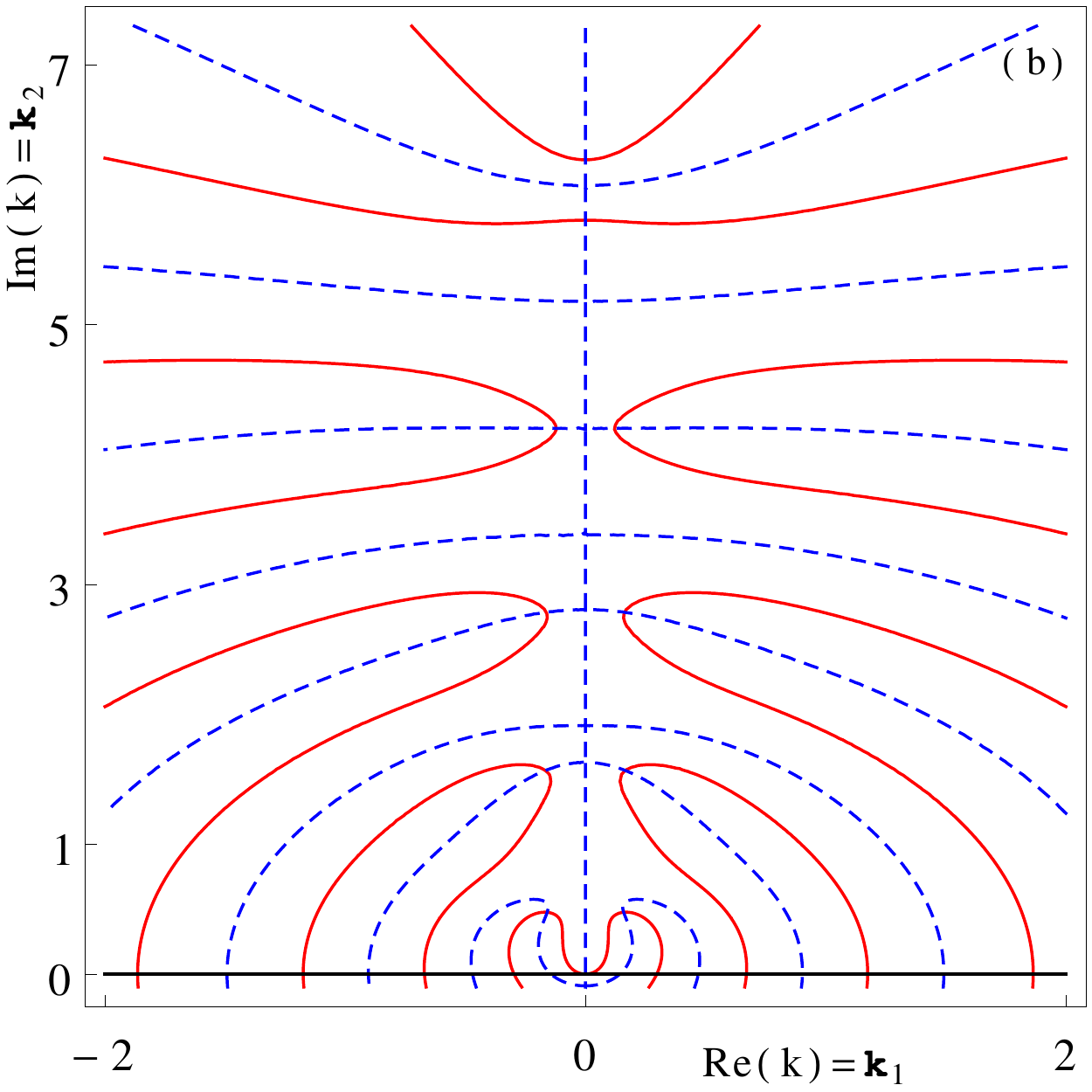}
	\caption{ The same as in Figs 1-3, for the exponential potential (17) $(a= 1 \AA$), see various rows in Table IV. (a): when $V_1{=}-60$ and $V_2{=}10$, there are 10 number of physical zeros of $F(k)$ as ten solid (red) curves are being intersected by the vertical dashed (blue) line. These roots on the $y-$axis represent bound states of the potential as $V_2$ is less than the fourth (smallest) EP which is $V_{EP4}=12.82$ (see Fig. 4(a)). There are no SDSS or CCEP in this case (row no. 18 in Table IV). In (b) we have $V_1=-60$ and $V_2=14$ that lies between two EPs $V_{EP1}=14.78$ and $V_{EP2}=13.63$ in this case there are two real negative energy bound states and rest are CCPEs (row no. 19 in Table IV).}
\end{figure}

First 17 rows in Table IV  for exponential potential are similar to those of Table I to III.
The distinctive feature of exponential model (17) is exhibited by row nos. $\{9, 14 , 16 \}$.
In these rows, the ${\#}$-tagged cases suggest slightly exceptional feature where $E_*$ is rough the upper bound to the real parts of CCEPs. These are 3 exceptions in overall 61 cases presented in four Tables for four different potential models.

Fig. 4 presents the evolution of negative real discrete  eigenvalues $E_n(V_2)$  for $V_1=-60, a=2$, see coalescing of five pairs of negative discrete eigenvalues at five exception points: $V_{EP1}{=} 12.82,~ V_{EP2}{=}12.96,~ V_{EP3}{=}13.17,~ V_{EP4}{=}13.63,$ $V_{EP5}{=}14.78.$ There are 10 pairs of eigenvalues when $V_2=10 <V_{EP1}$ (see row no. 18 of Table IV and Figs. 4(b)). These eigenvalues are also depicted as purely imaginary roots of $F(k)=0$ in upper $k$-plane in Fig. 5(a). When $V_{EP3}<V_2< V_{EP4}$ there are two real and four CCPEs see row no. 19 in Table IV and Fig. 5(b). When $V_2=83 >>V_{EP5}$, there exists an SDSS an ($E_*=8.551$) along with five CCPEs whose real parts are less than $E_*$ (see row no 20).

We confirm the phenomenon of CPA-Laser [3] at SDSS in all the four potential models presented here. This means that two port scattering matrix $S(E)$, $|\det(S(E))|=1$ which becomes indeterminate ($0/0$) at $E=E_*$ such that $\lim _{E \rightarrow E_*} |\det(S(E))| \rightarrow 1 [3].$
The localized CPTSSP (9,13), display unidirectional invisibility [14] at several energies of injection but other potentials (4,17) display it  for at most one energy.
\end{widetext}
\section{Conclusions}
We have shown that the poles of $t(k)$ or zeros of the transfer-matrix-element ${\cal M}_{22}(k)=1/t(k)$ of the type $\pm k_1+ik_2, k_2>0$ give rise to three types of discrete energy eigenvalues $(E_n=k^2)$. More importantly the complex-conjugate pairs of eigenvalues of a complex PT-symmetric scattering potential are yielded. Recently proposed splitting of the spectral singularity has also been confirmed. Alternatively, the zeros of ${\cal M}_{11}(k)$ of the type $\pm k_1 + i k_2, k_2<0$ can also yield  three types of discrete eigenvalues and they can  explain the phenomenon of splitting of spectral singularity. Here, it is the appearance of $\pm$ signs which is very crucial and which has been missed out earlier, also see Figs. 1-3,5 ; where $k-$poles of ${\cal M}_{22}(k)$ are symmetrically placed about $y-$axis. It will be well to point out that in general there can be algebraic or transcendental equations such as $f(x,i)=f(-x,-i)=0$ (PT-invariant), these equations have roots which are essentially of the type: $x =\pm a+ib$ where $a,b \in \Re$. In case of complex PT-symmetric scattering potentials the energy eigenvalue equations here and elsewhere are always of the type $F(k,i)=F(-k,-i)=0$ and hence three types of discrete energy eigenvalues.

Our extensive study of four exactly solvable and other numerically solved models reveals that in a CPTSSP if $V_2$  crosses an exceptional point $V_{EP}$ or a critical value $V_{*}$, a new CCPE is created. In the absence of EPs, if $V_2=V_{*m}$, the potential possesses one SDSS and $m$  number of CCPEs, otherwise these are more than $m$ number of CCPEs. This  gives rise to the  following conjectures about self-dual spectral singularity.

\noindent
$\bullet$ In CPTSSP, SDSS is essentially a phenomenon of parametric regime of broken PT-symmetry. In other words a CPTSSP  having real discrete eigenvalues cannot have an SDSS and vice-versa. Hence these two are mutually exclusive. See row no. 11 in Tables (I-IV) and Fig. 5 in this regard.\\
$\bullet$ A parametrically fixed CPTSSP can have at most one spectral singularity: none or one. No exception to this  exists  so far.\\
$\bullet$ If SDSS occurs, it is mostly the upper-bound  $(E_*>{\cal E}_l)$ and rarely the rough upper bound $(E_* \approx {\cal E}_l)$ to the real part of the CCPEs of the potential. Here the subscript $l$ means last. \\
$\bullet$ The SDSS at $E=E_*$ for $V_2=V_{*m}$ splits into a CCPE if $V_2$ is increased $(V_2>V_{*m})$. If for $V_2=V_{*m}$ there exist  one SDSS and $m$ number of CCPEs, then for $V_2=V_{*m} +\epsilon, \epsilon >0$ there is no SDSS but there will be $m+1$ CCPEs. For $V_2=V_{*m}-\epsilon$ again there is no SDSS and there will be $m$ CCPEs. The former is referred to as splitting of the SDSS. \\
$\bullet$ When the real part of a CPTSSP is a barrier, CCPE results essentially from splitting of the SDSS. Otherwise CCPE also results after coalescing of real discrete eigenvalues at an exceptional point.\\
$\bullet$ If a CPTSSP has Kato's exceptional points (EPs: $V_{EPn}=V_{EP1} <V_{EP2} <V_{EP3}<...<V_{EPl}$), the critical values of $V_2$ ($=V_{*m}$) for the SDSS are  larger than $V_{EPl}$.

We also find that localized potentials (square well and Dirac delta models) are uni-directionally invisible at multiple energies
whereas Scarf II and exponential potentials are so, for at most one energy of incidence. Bi-directional invisibility is rare, however, purely imaginary delta well model entails them in abundance.

We hope that the present study of three types of discrete eigenvalues, interesting results and conjectures about self-dual spectral singularity will open up a new direction of investigations in complex PT-symmetric potentials which have been giving rise to novel possibilities in the coherent injection at the optical mediums with equal gain and loss. 

\section*{Acknowlwdgements} We would like to thank Prof. V. V. Konotop for several email communications during our investigations. We also thank an anonymous Referee for a constructive suggestion.
\section*{References}
\begin{enumerate}
	\item V. V. Konotop and D. A. Zezyulin, Optics Letters {\bf 42} (2017).
	\item A. Mostafazadeh, Phys. Rev. Lett. {\bf 102} 220402 (2009). 
    \item S. Longhi, Phys. Rev. A {\bf 82} 031801R (2010).
    \item C. M. Bender and S. Boettcher, Phys. Rev. Lett. {\bf 80} 5243 (1998). 
    \item Z. Ahmed, Phys. Lett.  A {\bf 282}, 343 (2001); {\bf 287} 295 (2001). 
    \item Z. Ahmed,  Phys. Rev. A {\bf 64},  042716 (2001).
	\item V.V. Konotop, J. Yang and D.A. Zezyulin, Rev. Mod. Phys. {\bf 88} 035002 (2016).
	\item Z. Ahmed, Phys. Lett. A {\bf 377} 957 (2013).
	\item Z. Ahmed, D. Ghosh and S. Kumar,  Phys. Rev. A {\bf 97} 023828 (2018).
	\item Y. D. Chong, Li Ge, Hui Cao and A. D. Stone, Phys. Rev. Lett. {\bf 105} 053901 (2010).
   	\item Z. Ahmed, J. Phys. A: Math. Theor. {\bf 42}  472005 (2009).
   	 \item A. Khare, and U.P. Sukhatme, J. Phys. A: Math. Gen. {\bf 21}, L501 (1988).
    \item B. Bagchi and C. Quesne, J. Phys.  A: Math. Theor. {\bf 43} 305301 (2010).
    \item A. Mostafazadeh, Phys. Rev. A {\bf 87} 012103 (2013).
    \item Z. Ahmed, J. Phys. A: Gen. $\&$ Theor. {\bf 45} 032004 (2012).
   \item Z. Ahmed, J. A. Nathan and D. Ghosh, Phys. Lett. A {\bf 380}, 562 (2016).
   \item Z. Ahmed, Phys. Lett. A {\bf 286} 231 (2001).
  \item  H. Uncu and E. Demiralp, Phys. Lett. A {\bf 359} 190 (2006).
  \item A. Mostafazadeh and H. Mehri-Dehnavi, J. Phys. A Math. Theor.
   {\bf 42} 125303 (2009).  
     	\item Z. Ahmed, Phys. Lett. A {\bf 324} 152 (2004).	
   	\item Z. Ahmed, J. A. Nathan, D. Sharma and D. Ghosh, Springer Proceedings of Physics   {\bf 184} 1 (2016).

\end{enumerate}
 	\end{document}